\documentclass[runningheads]{llncs}
\usepackage{graphicx}
\usepackage{amssymb}
\usepackage{bbm}

\usepackage[english]{babel}
\usepackage[T1]{fontenc}
\usepackage[utf8]{inputenc}

\usepackage{booktabs} 
\usepackage{graphicx}
\usepackage{amsmath}
\graphicspath{{pics/}{figs/}}

\begin{document}
\title{Cascaded Deep Neural Networks for Retinal Layer Segmentation of Optical Coherence Tomography with Fluid Presence}
%
%
\author{Donghuan Lu\inst{1, 2} \and Morgan Heisler\inst{1}  \and
	Da Ma\inst{1} \and Setareh Dabiri\inst{1} \and
	Sieun Lee\inst{1} \and Gavin Weiguang Ding\inst{1} \and Marinko V. Sarunic\inst{1}
	\and Mirza Faisal Beg\inst{1}}

%
%
\institute{$^1$School of Engineering Science, Simon Fraser University\\
	$^2$Tencent YouTu Lab\\dla121@sfu.ca}
\maketitle              
\begin{abstract}
Optical coherence tomography (OCT) is a non-invasive imaging technology which can provide micrometer-resolution cross-sectional images of the inner structures of the eye. It is widely used for the diagnosis of ophthalmic diseases with retinal alteration, such as layer deformation and fluid accumulation. In this paper, a novel framework was proposed to segment retinal layers with fluid presence. The main contribution of this study is two folds: 1) we developed a cascaded network framework to incorporate the prior structural knowledge; 2) we proposed a novel deep neural network based on U-Net and fully convolutional network, termed LF-UNet. Cross validation experiments proved that the proposed LF-UNet has superior performance comparing with the state-of-the-art methods, and incorporating the relative distance map structural prior information could further improve the performance regardless the network.
\end{abstract}

\section{Introduction}
Optical coherence tomography (OCT) has been widely used to detect and monitor pathologies from retinal diseases. Anatomical and structural alteration measured from OCT images, such as layer thinning and fluid accumulation, are important signs for various types of retinal diseases \cite{fujimoto1995,joussen2010retinal}. However, manual segmentation of retinal layers and fluid is extremely time consuming, and suffers from inter-rater variability. Development of automatic segmentation tools can potentially help the physicians to achieve fast and accurate diagnosis.

The retinal layer segmentation methods can be categorized into two groups: 1) The mathematical model based methods construct the models using prior assumptions of image structure, such as global shape regularization \cite{rathke2014probabilistic} and graph \cite{lee2013comparative} based methods. 2) The pixel-wise classification based methods extracted pixel- or patch-wise features and feed to machine learning classifiers such as support vector machine (SVM) \cite{srinivasan2014automatic} and deep learning based neural network \cite{roy2017relaynet}. However, the performance of current available approaches are still behind the accuracy of human-rater's and new methods are needed for better segmentation accuracy.

In this study, a novel deep learning based framework is proposed to segment retinal layers with the presence of fluid. The major contributions of the study are: 1) Proposed a novel deep neural network, FL-UNet. Improved from the U-Net \cite{ronneberger2015u} and FCN \cite{long2015fully}, the proposed network outperformed state-of-the-art methods in the cross validation experiments. 2) Proposed a novel framework with cascading networks to incorporate prior structural knowledge in a specific designed form, i.e, relative distance map. By calculating the relative distance map based on the segmentation of the first network and use it as additional channel of input for the second network, the performance of proposed approach was further improved regardless the network used in the framework.

\section{Methods}
\label{sec:method}
\begin{figure}
\vspace{0.1cm}
\centerline{\includegraphics[width=11cm]{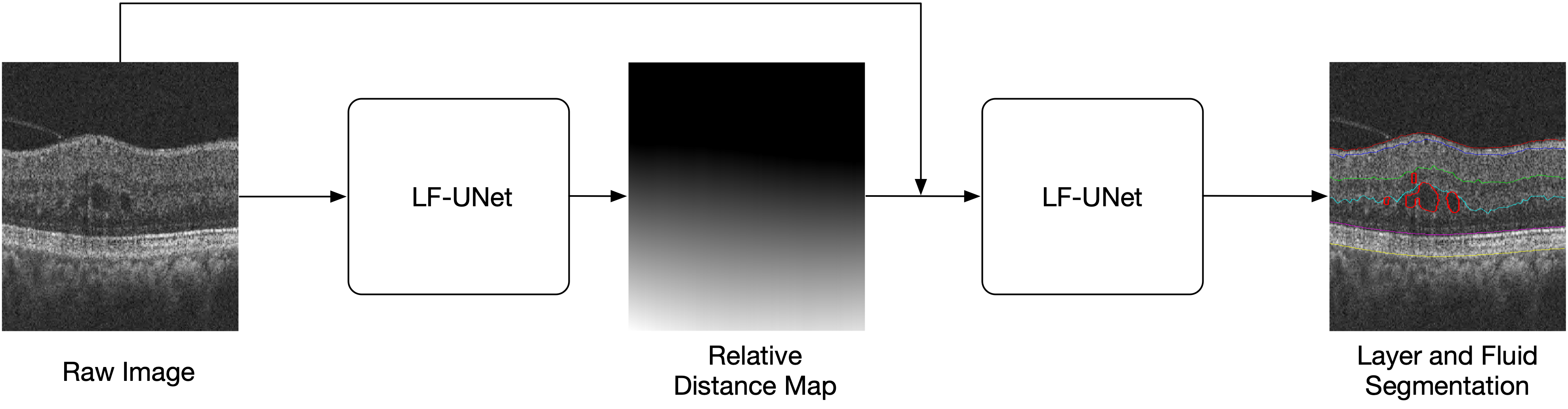}}
\caption{Flowchart of the proposed novel framework, comprising a cascade of two LF-UNet incorporating prior anatomical information, relative distance map within the retina, to achieve simutaneous layer and fluid segmentation.}
\label{fig:framework}
\end{figure}

Our framework for the segmentation of retinal layers and fluid consisted of two cascaded LF-UNet as displayed in Figure \ref{fig:framework}. First, the inner limiting membrane (ILM) and the Bruch's membrane (BM) were segmented by the first LF-UNet. Second, the relative distance map was calculated and used as an addition channel of input for the second LF-UNet to segment 6 retinal surfaces and fluid. A Random Forest classifier was trained in the last step to rule out false positive fluid regions as detailed in \cite{lu2019deep}. The final outcome is the segmentation of both retinal fluid and 6 layer surfaces, including the ILM, the posterior boundary of nerve fiber layer (NFL), the posterior boundary of inner plexiform layer (IPL), the posterior boundary of outer plexiform layer (OPL), the IS/OS junction, and the Brunch's membrane. 

\subsection{Materials}
\label{ssec:data}
The OCT images were acquired using a Zeiss Cirrus 5000 HD-OCT (Zeiss Meditec. Inc, Germany) which uses the OCT- microangiography complex algorithm (OMAG) with an A-scan rate of 68Khz. The 3x3mm pattern was used with a sampling rate of 245x245, which corresponds to a distance of $12.2 \mu m$ between scanning locations. A total of 4 B-scans were acquired at each location. The A-scan depth of the system is 2mm with an axial resolution of $5 \mu m$ and a transverse resolution of $15 \mu m$. A total of 58 3D volumes were used in this study, 25 of which are diabetic patients who mainly exhibited intraretinal fluid. We regarded all the fluid as a single class due to the limited available data.

\subsection{Network Architecture}
The network architecture of the proposed LF-UNet was a combination of the U-Net \cite{ronneberger2015u} and FCN \cite{long2015fully}. We labelled the pixels between two retinal layer boundaries as the same class for training, instead of determining the layer boundaries, effectively converting the problem of boundary detection into tissue segmentation. 5 retinal layer structures were segmented, here referred to as ILM-NFL, NFL-IPL, IPL-OPL, OPL-IOS and IOS-BM in the rest of this article. 

\begin{figure}
\vspace{0.1cm}
\centerline{\includegraphics[width=11cm]{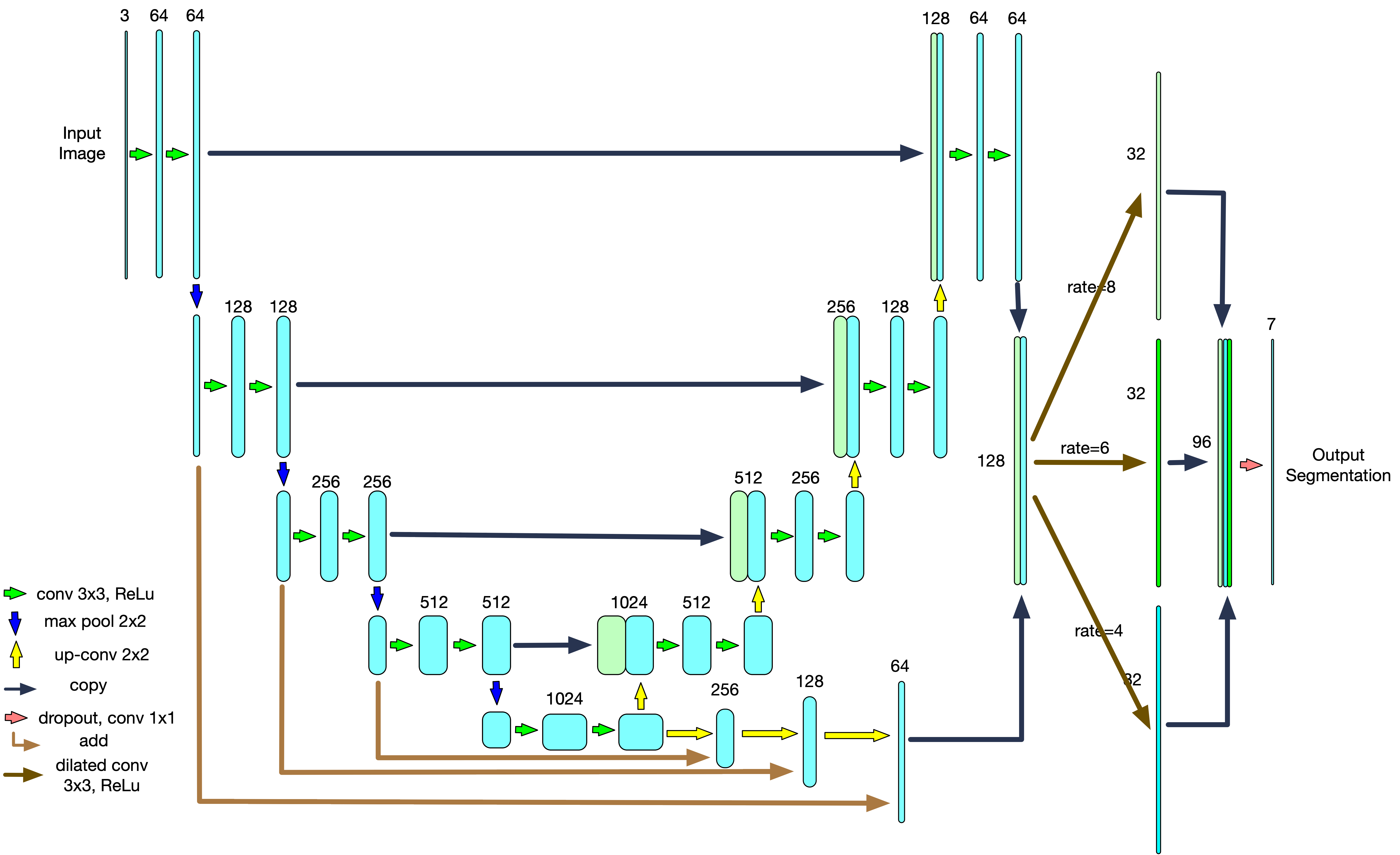}}
\caption{Network architecture of the LF-UNet, which is a combination of U-Net and Fully convolutional neural network. Each number above the cyan box represents the number of B-scans of the feature map.}
\label{fig:Cnetwork}
\end{figure}

As illustrated in Figure \ref{fig:Cnetwork}, the U-Net and the FCN shared the same contracting path (the downward path on the left side). It consisted of 4 blocks which contained two convolution layers with kernel size $3\times 3$ followed by an nonlinear activation function and a $2\times 2$ max pooling layer with stride 2. The number of feature maps in each block were 64, 128, 256 and 512, respectively. 

The expansive path was split into two parts: the U-Net part and the FCN part. In the U-Net part, the features extracted at each contracting block were concatenated with the features generated at the expansive block through short cut connections to provide high-resolution information. In the FCN part, the features of the contracting block and the expansive block with same resolution were added up as the input for the next block. The U-Net had good performance on the estimation of a coarse segmentation; however, it did not work well at extracting the boundaries. On the other hand, the FCN was reliable for the extraction of boundaries, while a large number of data samples was required for its training. The combination of these two networks harnesses their individual strengths, which leads to better segmentation. $2\times 2$ up-convolution layer were used in both parts after convolutional layers.

The feature maps of the last convolutional layers in both parts were concatenated and fed to the three parallel dilated convolution layers \cite{yu2015multi} followed by a single layer 1x1 convolutional network to predict the segmentation of different layers and the fluid. As some retinal layers occupied a large area, we used dilated convolutional layers instead of normal convolutional layers to increase the receptive field, enabling us to harness enough information from nearby layers. Using a large receptive field with normal convolutional layers would result in many more parameters thereby requiring more computational resources and potentially leading to overfitting. Conversely, dilated convolutional layers can enlarge the receptive field without increasing the number of parameters by skipping some units during convolution. All the activation functions used in the hidden layers were Rectified linear units (Relu), while the Softmax function was used for the output layer to map the output probability to $(0,1)$.


\subsection{Relative Distance Map}
The segmentation of the ILM and the BM was relatively easy due to their strong contrast compared with other layers and the background, while the segmentation of the remaining layers was more challenging due to their relative similar intensity patterns, especially for the posterior boundary of NFL, the posterior boundary of IPL and the posterior boundary of OPL, as shown in the right image of Figure \ref{fig:framework}. An important feature to determine the layer label of each pixel is its location in the retina. However, such features can hardly be captured by the network itself since the convolution kernel can only capture the information of nearby pixels. Relative distance maps \cite{lu2019deep} have proven to be a useful feature for the segmentation of multiclass fluid in the retina, and its potential usage in layer segmentation is worth exploring. For pixel $(x,y)$ in a B-scan, its intensity in the relative distance map is defined as:
\begin{equation} \label{eq:location}
I(x,y)=\frac{y-Y_{1}(x)}{Y_{1}(x)-Y_{2}(x)} \quad
\end{equation}
where $Y_{1}(x)$ and $Y_{2}(x)$ represent the $y$-coordinate of ILM and BM segmented by the first LF-UNet, respectively. 

The relative distance map was concatenated to the B-scans as an additional channel of input for the second LF-UNet. Because the relative distance of background pixels above ILM was less than 0, while the relative distance of background pixels below ILM was larger than 1, they were labeled differently to avoid the confusion of network, resulting 8 mutually exclusive classes, background above ILM, ILM-NFL, NFL-IPL, IPL-OPL, OPL-IOS, IOS-BM, background below BM, and the fluid.

\subsection{Loss function}
The network was trained end-to-end with a loss function which consisted of two parts: the weighted Dice loss and the weighted logistic loss \cite{roy2017relaynet}.
The weighted Dice loss was defined as:
\begin{equation} \label{eq:diceloss}
Loss_{Dice}=1-\frac{2\sum_{x\in\Omega}\omega_lp_l(x)g_l(x)}{\sum_{x\in\Omega}p_l^2(x)+\sum_{x\in\Omega}g_l^2(x)}\
\end{equation}
where $\Omega$ represents the retinal region, $g_l(x)$ is the ground truth, $p_l(x)$ is the estimated probability of pixel $x$ belongs class $l$. $\omega_l$ is the weight associated with the number of pixels in different classes to resolve the imbalance among different layer regions and the fluid.

The weighted logistic loss is defined as below:
\begin{equation} \label{eq:weightloss}
Loss_{log}=-\sum_{x\in\Omega}\omega(x)g_l(x)log(p_l(x))
\end{equation}
where $\omega$ is the weight associate with each pixel $x$. 

In order to make the network more sensitive to boundary and retinal regions, the weight is designed as:
\begin{equation} \label{eq:weightloss}
\omega(x)=1+\omega_1I(|{\bigtriangledown}l(x)|>0)+\omega_2I(l(x)=L)
\end{equation}
where $I$ represents an indicator function and $\bigtriangledown$ is the gradient operator. It is worth mentioning that $l$ is the label of pixel $x$ instead of its intensity, therefore a pixel with $|{\bigtriangledown}l(x)|>0$ must be a boundary pixel based on its ground truth segmentation. $L$ represents the entire retina, including fluid and 5 layer regions. $\omega_1$ and $\omega_2$ were set as 10 and 5, respectively. 

The overall loss function was defined as:
\begin{equation} \label{eq:weightloss}
Loss_{log}=\lambda_1Loss_{Dice}+\lambda_2Loss_{log}
\end{equation}
where $\lambda_1$ and $\lambda_1$ were set as 0.5 and 1, respectively.

\subsection{Optimization}
Due to the limitation of GPU memory, the segmentation was performed on each 2D B-scan instead of the whole 3D volume. The adjacent B-scans (one before and one after the B-scan to be segmented) were also used for segmentation considering the consistency of the retinal layers and fluid, resulting a 500x245x3 matrix in the input of the first LF-UNet.

Two strategies were used during the training stage to make the proposed network less prone to overfitting. First, a dropout layer was inserted between the dilated convolutional layers and the $1\times 1$ convolutional layer. The dropout ratio was set as 0.5 which means only half of the units were randomly retained to feed features to the last convolutional layer in the training stage. By avoiding training all units on every sample, this regularization technique not only reduced the chances of overfitting by preventing complex co-adaptations on the training data, but also reduced the amount of computation and improved training speed. Secondly, data augmentation was applied to create more training samples to improve the robustness and invariance properties of the network. Three types of image transformations were applied to augment the data - flip, rotation and scaling -  with rotation degree set as $-25^\circ$ to $25^\circ$ and the maximum scaling ratio set as $0.5$.

Batch size was set as 3 due to the GPU memory limitation. The weight parameters for each layer were initialized with a uniform distribution while all bias started with 0 \cite{glorot2010understanding}. Adaptive Moment Estimation (adam) optimizer was used for training with a fixed learning rate of $10^{-5}$ and the optimization was stopped if the training accuracy ceased to increase after 5 epochs. 

\section{Experiments and Results}
The deep neural network was built with Tensorflow \cite{tensorflow2015-whitepaper}, an open source deep learning toolbox provided by Google. All the experiments were run on NVIDIA P100-PCIE GPUs. To validate the ability of proposed framework, a 10-fold cross validation was performed on the 58 $3\times3mm$ volumes. To avoid the bias caused by using B-scans of the same volume in both training and testing, the volumes were divided into the training set which contained the B-scans of 52-53 volumes, and the testing set which contained the B-scans from the rest of the volumes in each cross validation experiment. The segmentation performance was evaluated using the Dice index for each B-scan. Performance for each layer and the fluid were measured separately, and the B-scans which did not contain fluid were discarded when measuring the performance of the fluid segmentation. 

\begin{table}[ht]
\scriptsize
\centering
\caption{Dice index of different networks and inputs. 1U-Net-1Bscan represents only a single U-Net which used a single B-scan as input, while 2LF-UNet-3Bscan means 2 concatenated LF-UNets were used for segmentation with 3 adjacent B-scans as input, and the relative distance map calculated from the first LF-UNet was used as the additional channel of input for the second LF-UNet. Column 2 to 6 represents the segmentation accuracy for layers from top to bottom. Column 7 and 8 represents the Dice index for fluid before and after random forest classification. Noticing the 2LF-UNet-3Bscan has the best performance regarding the segmentation of fluid and most layers.}
\begin{tabular}{@{}|c|c|c|c|c|c|c|c|@{}}
\toprule
               & ILM-NFL         & NFL-IPL         & IPL-OPL         & OPL-IOS         & IOS-BM          & Fluid           & RF-Fluid        \\ \midrule
1U-Net-1Bscan & 0.8679          & 0.9315          & 0.9033          & 0.9153          & 0.9308          & 0.5011          &0.3736           \\ \midrule
2U-Net-1Bscan & 0.8807          & 0.9454          & 0.9243          & 0.9411          & 0.9384          & 0.4871          &0.4293          \\ \midrule
1U-Net-3Bscan & 0.8910          & 0.9451          & 0.9180          & 0.9271          & 0.9329          & 0.4761          &0.3839           \\ \midrule
2U-Net-3Bscan & 0.9019          & 0.9526          & 0.9316          & 0.9458          & 0.9401          & 0.5066          &0.4770           \\ \midrule
1RelayNet-1Bscan & 0.9032          & 0.9472          & 0.9208          & 0.9424          & 0.9432          & 0.4177          & 0.4079          \\ \midrule
2RelayNet-1Bscan & 0.9075          & 0.9500          & 0.9261          & 0.9472          & 0.9452          & 0.4676          & 0.4955          \\ \midrule
1RelayNet-3Bscan & 0.9236          & 0.9580          & 0.9355          & 0.9472          & 0.9451          & 0.4313          & 0.4299          \\ \midrule
2RelayNet-3Bscan & 0.9255          & 0.9593          & 0.9379          & 0.9517          & 0.9471          & 0.4471          & 0.3977          \\ \midrule
1LF-UNet-1Bscan & 0.9100          & 0.9531          & 0.9278          & 0.9466          & 0.9439          & 0.5014          & 0.4922          \\ \midrule
2LF-UNet-1Bscan & 0.9063          & 0.9507          & 0.9281          & 0.9484          & 0.9446          & \textbf{0.5132}          & 0.5661          \\ \midrule
1LF-UNet-3Bscan &\textbf{0.9283}         & 0.9610          & 0.9388          & 0.9509          & 0.9459          & 0.4674         & 0.4624          \\ \midrule
2LF-UNet-3Bscan  &0.9278 & \textbf{0.9612} & \textbf{0.9409} & \textbf{0.9526} & \textbf{0.9466} & 0.4985  & \textbf{0.5837} \\ \bottomrule
\end{tabular}
\label{table:crossvalidation}
\end{table}

Table \ref{table:crossvalidation} shows the comparison of the performance between the proposed framework and two state-of-the-art methods: the U-Net \cite{ronneberger2015u} and the RelayNet \cite{roy2017relaynet}. Two kinds of input, single B-scans and 3 consecutive B-scans, were tested with or without a relative distance map. The evaluation results showed that, 1) Under the same condition, the proposed LF-UNet had better performance comparing with U-Net and RelayNetwork regarding the segmentation of both the retinal layers and fluid; 2) Using the two adjacent B-scans as addition channels of the input could improve the segmentation accuracy regardless of the network architecture; 3) Under most circumstances, cascading networks to incorporate prior structural knowledge of retina could further improve the performance, suggesting the relative distance is a useful feature not only for multiclass fluid segmentation, but also for retinal layer segmentation.

Comparing with the layer segmentations, the fluid segmentation showed inferior performance with all three network architectures. It is due to insufficient number of available training samples that contains fluid as well as the similarity between the signal of the fluid and shadow artifact. It is worth mentioning that although the value of Dice index of the fluid segmentation was lower than the layer segmentation, it showed at least 0.08 improvement comparing with the other two networks, suggesting a better discriminant ability of the proposed network. For some experiments with U-Net or RelayNet, the dice index of fluid did not increase after random forest classification. It is because random forest classification is better at rule out false positive regions, but unable to identify the false negative regions - the fluid region missed by the network. Therefore, for networks that are not very sensitive to fluid, the Dice index may even decrease after random forest classification. 

\section{Conclusion}
In this paper, we have proposed a novel framework to automatically segment retinal layers as well as fluid in OCT images. A novel deep neural network, LF-UNet, was proposed and cross validation experiments proved that the proposed network outperformed state-of-the-art methods. Further experiments showed that by cascading two networks, incorporating structural prior knowledge using the relative distance map derived from the first network could improve the segmentation performance regardless of the network. Due to the limited number of data samples with fluid, the segmentation result of fluid are comparatively not as high as layer regions. As more data is accumulated in the future, it is likely that further improvements of fluid segmentation can be achieved.


%
%
\bibliographystyle{splncs04}
%
\bibliography{main}
\end{document}